\documentclass[twocolumn,aps,prc,superscriptaddress,nofootinbib,floatfix,longbibliography]{revtex4}
\usepackage{url}
\usepackage{cancel}
\usepackage[colorlinks,linkcolor=blue,citecolor=blue,filecolor=black,urlcolor=blue]{hyperref}
\usepackage{epsfig,graphics}
\usepackage{graphicx}
\usepackage{dcolumn}
\usepackage{bm}
\usepackage{amssymb}
\usepackage{amsmath}
\usepackage{multirow}
\usepackage{float}
\usepackage{harpoon}
\usepackage{MnSymbol}
\usepackage{appendix}
\usepackage{color}
\usepackage{hyperref}
\usepackage{cleveref}

\newcommand{\nch} {N_{\mathrm{ch}}}
\newcommand{\snn}{\mbox{$\sqrt{s_{\mathrm{NN}}}$}}
\newcommand{\pT} {p_{\mathrm{T}}}
\newcommand{\lr}[1]{\left\langle #1\right\rangle}

\newcommand{\npart}{N_{\mathrm{part}}}
\newcommand{\pa} {p_{\mathrm{T1}}}
\newcommand{\pb} {p_{\mathrm{T2}}}
\newcommand{\ea} {\eta_{\mathrm{1}}}
\newcommand{\eb} {\eta_{\mathrm{2}}}
\newcommand{\pTref} {p_{\mathrm{T}}^{\mathrm{ref}}}
\newcommand{\pTc} {p_{\mathrm{T}}^{0}}
\graphicspath{{logos/}{figs/}}

\begin{document}
\title{The shape of differential radial flow $v_0(p_T)$, not its zero-crossing, carries physical information}
\newcommand{\bnl}{Physics Department, Brookhaven National Laboratory, Upton, NY 11976, USA}
\newcommand{\sbu}{Department of Chemistry, Stony Brook University, Stony Brook, NY 11794, USA}
\author{Somadutta Bhatta}\affiliation{\sbu}
\author{Aman Dimri}\affiliation{\sbu}
\author{Jiangyong Jia}\email[Correspond to\ ]{jiangyong.jia@stonybrook.edu}\affiliation{\sbu}\affiliation{\bnl}
\date{\today}
\begin{abstract}
Radial flow, a key collective phenomenon in heavy-ion collisions, manifests itself through event-by-event fluctuations of transverse-momentum ($p_{\mathrm{T}}$) spectra. The $p_{\mathrm{T}}$-differential radial flow observable, $v_0(p_{\mathrm{T}})$, was introduced to quantify local spectral-shape fluctuations, but it is unavoidably influenced by global multiplicity fluctuations. Using the HIJING model, we show that different event-activity definitions for centrality classification and different spectral normalization schemes generate a constant vertical offset in $v_0(p_{\mathrm{T}})$ without altering its shape. This offset reflects the impact of residual volume/centrality fluctuations rather than genuine dynamical radial flow fluctuations. Accordingly, only the shape of $v_0(p_{\mathrm{T}})$, or equivalently its derivative $dv_0(p_{\mathrm{T}})/dp_{\mathrm{T}}$, carries physical information about radial-flow dynamics; its zero crossing does not. Practical implications include the need to vertically align measurements from different experiments before comparison, thereby removing normalization ambiguities when constraining QGP properties.
\end{abstract}
\maketitle

{\bf Introduction.} 
Radial and anisotropic flows are hallmark observables of collective behavior in the quark-gluon plasma (QGP) formed in high-energy heavy-ion collisions~\cite{Heinz:2024jwu}. While anisotropic flow ($v_n$, $n=1,2,...$) has been extensively studied, radial flow ($v_0$)---the isotropic collective expansion driven by pressure gradients---remains comparatively less explored. Event-by-event (EbE) fluctuations in the initial collision geometry and local energy densities produce corresponding fluctuations in radial flow, which manifest as variations in the transverse momentum ($\pT$) spectra, $N(\pT) = dN/d\pT$~\cite{Bozek:2012fw,Samanta:2023amp}. These radial flow fluctuations encode valuable information about fundamental QGP properties, including its equation of state and transport coefficients such as shear and bulk viscosities~\cite{Heinz:2013th,CMS:2024sgx,ATLAS:2024jvf}.

Historically, radial flow studies focused on integral quantities such as the average transverse momentum $[\pT]_R$ or particle multiplicity $N_R$ within a momentum range $R$~\footnote{In this paper, we use ``[]'' to denote average of a quantity in a given event, whereas $\lr{}$ denotes average over an event ensemble.}. These studies quantify radial flow fluctuations through multi-particle cumulants, $c_0\{k\}$ and $v_0\{k\}$ ($k=1,2,...$), formulated analogously to anisotropic flow:
\begin{align}\nonumber
&c_{0,x}\{1\} = v_{0,x}\{1\}   = \lr{x} \;, \\\nonumber
&c_{0,x}\{2\} = v_{0,x}\{2\}^2 = \lr{(\delta x)^2}/\lr{x}^2 \;,\\\nonumber
&c_{0,x}\{3\} = v_{0,x}\{3\}^3 = \lr{(\delta x)^3}/\lr{x}^3 \;,\\\nonumber
&c_{0,x}\{4\} = v_{0,x}\{4\}^4 = \frac{\lr{(\delta x)^4}-3\lr{(\delta x)^2}^2}{\lr{x}^4}\;,\\\label{eq:1}
&...
\end{align}
where $x=N_R$ or $[\pT]_R$, and $\delta x = x-\lr{x}$. For brevity, we denote $v_{0}\{k\}$ when $x=N_R$ and $v_{0,p}\{k\}$ when $x=[\pT]_R$. Experiments have measured the mean and variance of $[\pT]$ (i.e., $v_{0,p}\{1\}$ and $v_{0,p}\{2\}$), though measurements of higher-order cumulants have only recently begun to emerge~\cite{NA49:1999inh,ALICE:2014gvd,Adam:2019rsf,ALICE:2023tej,ATLAS:2024jvf,STAR:2024wgy}.  

The integrated cumulants $v_0\{k\}$ and $v_{0,p}\{2\}$ characterize multiplicity and $[\pT]$ fluctuations respectively, but do not resolve how the spectral-shape fluctuation varies with $\pT$. Just as for anisotropic flow, where the integrated $v_n\{2\}^2=c_n\{2\}$ is generalized to the $\pT$-differential $v_n(\pT)$ via the factorization $c_n(\pa,\pb) = v_n(\pa)v_n(\pb)$, one defines $v_0(\pT)$ through the analogous relation $c_0(\pa,\pb) = v_0(\pa)v_0(\pb)$. Very recently, measurements have advanced beyond these integral observables by studying two-particle radial flow fluctuations differentially in $\pT$ or pseudorapidity $\eta$~\cite{ATLAS:2025ztg,ALICE:2025iud}.

Through rigorous tests of factorization relations~\cite{ATLAS:2025ztg}, such as $c_{0}\{2\}(\pa,\pb)=v_{0}\{2\}(\pa)v_0\{2\}(\pb)$ and $c_{0}\{2\}(\ea,\eb)=v_{0}\{2\}(\ea)v_0\{2\}(\eb)$, these measurements confirmed the collective and long-range nature of radial flow. The measured $v_0(\pT)$~\cite{ATLAS:2025ztg} displays characteristic behavior: it starts negative at low $\pT$, increases and crosses zero near $\pT=\pTc \approx \lr{[\pT]}$ around 1~GeV, then peaks around 3--4~GeV before decreasing at higher $\pT$. Furthermore, the $\pT$-dependent shape of $v_{0}\{2\}(\pT)$ was identified as particularly sensitive to bulk viscosity~\cite{Parida:2024ckk,ATLAS:2025ztg}, offering a new avenue for constraining QGP transport properties. This paper focuses exclusively on $v_{0}\{2\}(\pT)$, hereafter denoted simply as $v_{0}(\pT)$.

\begin{figure}[h!]
    \centering
    \includegraphics[width=0.75\linewidth]{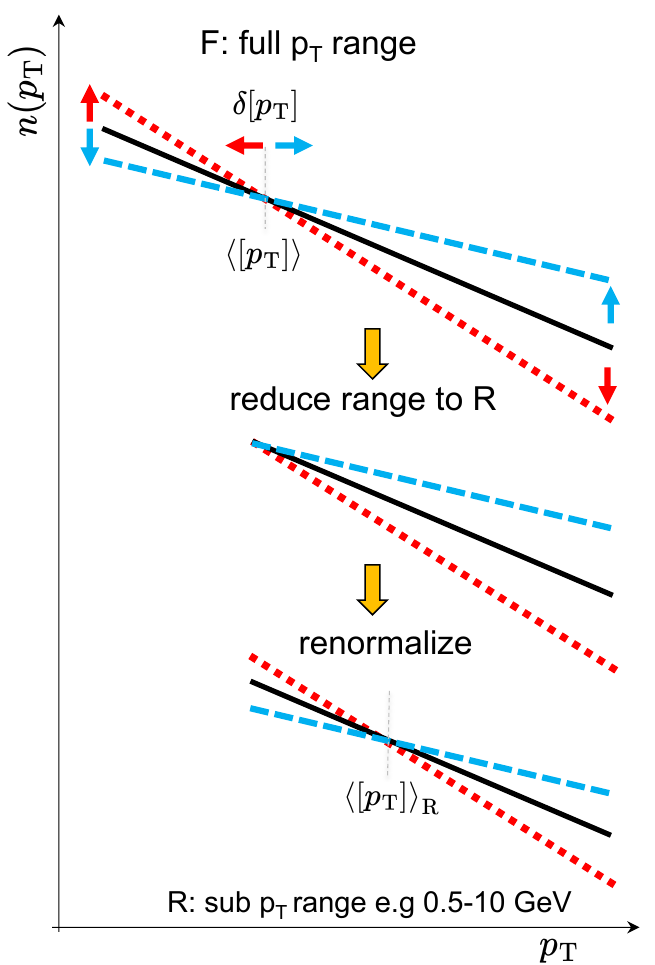}
    \caption{Top: Schematic illustration showing how radial flow fluctuations create correlations between the EbE $\pT$-differential yield $n(\pT)$ and the EbE average transverse momentum $[\pT]$. The blue curve represents an event with larger-than-average radial flow, the red curve represents smaller-than-average radial flow, and the black curve represents the ensemble average. The zero-crossing point of $v_0(\pT)$, approximately at $\langle [\pT] \rangle$, depends on the normalization range chosen to obtain $n(\pT)$. Bottom: Fractional spectra obtained over a wider range $F$ and its subrange $R$ are normalized differently. The spectral fluctuation in range $R$ can be decomposed into a global multiplicity fluctuation term and a reduced spectral shape fluctuation term (Eq.~\eqref{eq:c}).}
\vspace*{-0.3cm}
\label{fig:1}
\end{figure}

The radial flow in each event is encoded in the $\pT$-dependent yield, $N(\pT)= N n(\pT)$, where $N = \int N(\pT) d\pT$ is the global multiplicity and $n(\pT)$ is the normalized fractional spectrum satisfying $\int n(\pT) d\pT = 1$. Events with stronger (weaker) radial flow exhibit flatter (steeper) $N(\pT)$ distributions (see Fig.~\ref{fig:1} top). The EbE fluctuations of $N(\pT)$ can be decomposed as~\footnote{This decomposition is consistent with principal component analysis approaches~\cite{Mazeliauskas:2015efa,Gardim:2020fxx}.}:
\small{\begin{align}\nonumber
\delta N(\pT) &=  \lr{n(\pT)}\delta N + \lr{N} \delta n(\pT) +\delta N \delta n(\pT)\;.\\\label{eq:2}
              &\approx  \lr{n(\pT)}\delta N + \lr{N} \delta n(\pT)\;,
\end{align}}\normalsize
where the $\delta N$ term reflects centrality-dependent global multiplicity fluctuations, while the $\delta n(\pT)$ term captures genuine radial-flow fluctuations.

However, a fundamental ambiguity remains in separating $\delta N$ from $\delta n(\pT)$ when studying spectral fluctuations. We demonstrate that $v_0(\pT)$ is meaningful only up to an arbitrary constant reflecting irreducible centrality fluctuations. Crucially, this offset does not affect integrated observables such as $[\pT]$ fluctuations, confirming that only the shape or derivative $dv_0(\pT)/d\pT$ carries physical information about radial flow.  These findings have important practical implications: $v_0(\pT)$ measurements from different experiments or theoretical calculations must be vertically shifted to have the same zero-crossing to remove normalization ambiguities, before meaningful comparisons can be made.  

{\bf Impact of global multiplicity and spectral shape fluctuations on $v_0(\pT)$.} 
Radial flow information is inferred from the correlation between $n(\pT)$ and $[\pT]$, which can be expressed in a factorizable form~\cite{Schenke:2020uqq}:
\begin{align}\label{eq:3}
\frac{\lr{\delta n(\pT) \delta [\pT]}}{\lr{n(\pT)}\lr{[\pT]}} = v_0(\pT) v_{0,p}\;,
\end{align} 
where $v_{0,p} = \sqrt{\lr{(\delta [\pT])^2}_A}/\lr{[\pT]}_A$ is measured within a reference $\pT$ range $\pTref\in A$, and $v_0\{1\}(\pT) = \lr{n(\pT)}$ according to Eq.~\eqref{eq:1}. Recent ATLAS measurements indicate that $v_0(\pT)$ remains approximately independent of the choice of $\pTref$~\cite{ATLAS:2025ztg}, consistent with the collective nature of radial flow.

The connection between $v_0(\pT)$ and local spectral shape fluctuations can be explicitly formulated as:
\begin{align}\label{eq:4}
v_0(\pT) = \rho(n(\pT),[\pT]) \frac{\sqrt{\lr{(\delta n(\pT))^2}}}{\lr{n(\pT)}}\;,
\end{align} 
where $\rho(x,y)=\lr{\delta x \delta y}/\sqrt{\lr{(\delta x)^2}\lr{(\delta y)^2}} \in [-1,1]$ is the Pearson correlation coefficient between $x$ and $y$. Since $|v_0(\pT)|\leq \sqrt{\lr{(\delta n(\pT))^2}}/\lr{n(\pT)}$, $v_0(\pT)$ captures only the component of spectral fluctuation that is correlated with $[\pT]$. 

Having separated the spectral fluctuations into shape and multiplicity components (Eq.~\eqref{eq:2}), we can now state the central consequence of this decomposition.  Consider an alternative definition of $\pT$-differential radial flow using the unnormalized spectrum $N(\pT)$ instead of $n(\pT)$:
\begin{align}\label{eq:5}
\frac{\lr{\delta N(\pT) \delta [\pT]}}{\lr{N(\pT)}\lr{[\pT]}} = v_0'(\pT) v_{0,p}\;.
\end{align} 
Substituting Eq.~\eqref{eq:2} and using the approximation $\lr{N(\pT)} = \lr{N n(\pT)} = \lr{N}\lr{n(\pT)} (1+\tau)$, where the correction term 
\begin{align}\label{eq:5b}
\tau(\pT)=\lr{\delta N \delta n(\pT)}/(\lr{N}\lr{n(\pT)})
\end{align} is small for sufficiently narrow centrality bins, we obtain:
\begin{align}\label{eq:6}
\Delta v_0 &\equiv v_0'(\pT) - v_0(\pT) = \rho(N, [\pT]) \frac{\sqrt{\lr{(\delta N)^2}} }{\lr{N}}\;.
\end{align}
This result demonstrates that the two definitions differ by a constant. The sign and magnitude of this offset depend on how the global multiplicity $N$ correlates with $[\pT]$, which is sensitive to the specific centrality definition employed, for example, whether centrality is based on forward calorimeters or midrapidity charged particles~\cite{ATLAS:2025ztg}. In other words, this offset reflects the impact of volume or centrality fluctuations rather than genuine dynamical radial flow fluctuations.

{\bf Offset in $v_0(\pT)$ induced by the $\pT$-range choice.} 
While using the normalized spectrum $n(\pT)$ appears useful to isolate spectral shape fluctuations from global multiplicity effects, this separation is non-unique. The key issue is that experimental measurements are always restricted to a finite $\pT$ range $R$. The fractional spectrum defined within this range, $n_R(\pT) = N(\pT)/N_R$ where $N_R = \int_R N(\pT) d\pT$, differs from the ideal full-range spectrum by a rescaling factor: $n_R(\pT) = n(\pT) N/N_R$ (see Fig.~\ref{fig:1} bottom). Even if the total multiplicity $N$ were fixed, the multiplicity within the restricted range, $N_R$, would still fluctuate statistically due to the stochastic nature of particle production. 

This residual fluctuation introduces an additional layer of ambiguity similar to Eq.~\eqref{eq:2}, which can be decomposed as:
\begin{align}\label{eq:c}
\delta N(\pT) \approx \delta N_R \lr{n_R(\pT)} + \lr{N_R} \delta n_R(\pT)\;,
\end{align}
where $n_R(\pT) = N(\pT)/N_R$. Hence, applying the same logic, the $v_0(\pT)$ obtained in the full range and range $R$ should also differ by a 'range-induced' offset  (see Fig.~\ref{fig:1} bottom), whose value can be estimated as follows.

Assuming particles are independently sampled from the underlying momentum distribution, $N_R$ follows a binomial distribution with sampling probability $\epsilon = \lr{N_R}/\lr{N}$ (we ignore the possible $\pT$-dependent acceptance effect in the real experiment). The variance of $N_R$ at fixed $N$ is given by $\lr{(\delta N_R)^2}_N = N(1-\epsilon)\epsilon \approx \lr{N_R}(1-\epsilon)$. Applying Eq.~\eqref{eq:6}, the $v_0(\pT)$ calculated for $R$, denoted $v_{0,R}(\pT)$, differs from $v_0(\pT)$ defined in the full range by:
\small{\begin{align}\label{eq:7aa}
\Delta v_{0,R} &= v_{0,R}(\pT) - v_0(\pT)\\\label{eq:7a}
 &\approx \!\rho(N, [\pT])\!\frac{\sqrt{\lr{(\delta N)^2}} }{\lr{N}}\! -\!\rho(N_R, [\pT])\!\frac{\sqrt{\lr{(\delta N_R)^2}} }{\lr{N_R}}\\\label{eq:7b}
 &\approx  -\left(\rho(N_R, [\pT]) \frac{\sqrt{\lr{(\delta N_R)^2}} }{\lr{N_R}}\right)_N\\\label{eq:7c}
 & =-\left(\rho(N_R, [\pT])  \sqrt{\frac{{(1-\epsilon)}}{N_R}}\right)_N\;,
\end{align}}\normalsize 
where the subscript $N$ denotes evaluation at fixed $N$. The approximation from Eq.~\eqref{eq:7a} to Eq.~\eqref{eq:7b} holds because changes in $N$ induce proportional changes in $N_R$, such that the first term in Eq.~\eqref{eq:7a} becomes negligible.

This vertical offset translates into a horizontal shift in the zero-crossing point of $v_0(\pT)$ from $\pTc\approx\lr{[\pT]}$ to $\pTc\approx \lr{[\pT]}_R$. Importantly, this shift does not affect the factorization property (Eq.~\eqref{eq:3}), which remains valid regardless of the $\pTref$ range used to calculate $v_{0,p}$. As demonstrated in the Appendix, this offset also leaves integrated observables such as $[\pT]$ fluctuations unchanged due to sum rule constraints. Therefore, only the $\pT$-dependent variation of $v_0(\pT)$ contains meaningful information about collective radial flow fluctuations. 

{\bf Model setup. } To investigate the interplay between global multiplicity and spectral shape fluctuations, we employ the HIJING model~\cite{Gyulassy:1994ew}, which  treats heavy-ion collisions largely as a superposition of independent $pp$ collisions. The absence of genuine collective radial flow in HIJING, combined with the presence of high-$\pT$ particles from hard-scattered jets, provides an ideal framework for studying the non-flow baseline of $v_0(\pT)$ and isolating the effects of multiplicity fluctuations. However, HIJING cannot validate the sensitivity of $v_0(\pT)$ to genuine viscous or equation-of-state effects of QGP, since it contains no hydrodynamic expansion. Establishing the physical sensitivity of $v_0(\pT)$ to transport coefficients requires comparison with hydrodynamic models, which is beyond the scope of the present study.

We generate Pb+Pb collisions at $\snn=5.02$ TeV with a minimum hard-scattering scale of 2~GeV (parameter HIPR1(10)=2) to match LHC kinematics. Charged particles with $|\eta| < 2.5$ and $\pT<10$~GeV are selected, consistent with typical experimental acceptances. Following the two-subevent method employed by ATLAS~\cite{ATLAS:2025ztg} and STAR~\cite{STAR:2024wgy}, we divide particles into two pseudorapidity regions: $-2.5<\eta_a<-\eta_{\mathrm{gap}}/2$ and $\eta_{\mathrm{gap}}/2<\eta_b<2.5$. We use two pseudorapidity gap sizes, $\eta_{\mathrm{gap}}= 0$ and 3, to evaluate the effects of short-range non-flow correlations and to assess possible longitudinal multiplicity decorrelation effects.

The normalization ambiguities are studied by varying three key ingredients in our analysis: (1) We calculate $[\pT]$ using three different $\pTref$ ranges denoted by ``$A$'': 0.5--2, 0.5--5, and 1--5~GeV. This variation allows us to study how non-flow correlations, which are more prominent at high $\pT$, affect the extracted $v_0(\pT)$. (2) We define the fractional spectrum $n(\pT)$ over three $\pT$ ranges: 0--10~GeV (full range), 0.5--10~GeV, and 1--10~GeV (subranges denoted by ``$R$''). (3) We classify events using $\nch$ in various $\eta$ ranges (listed in Table~\ref{tab:1}) or the number of participating nucleons $\npart$ from the Glauber model. Both the second and third variations probe the impact of residual multiplicity fluctuations.

\vspace*{-0.1cm}
\begin{table}[!h]
\begin{tabular}{ c|c |c }
$n(\pT)$ or $N(\pT)$     & $\pTref$  (GeV) & Event activity \\\\\hline
 0--10 GeV, $|\eta|<2.5$ & 0.5--2  & $\nch$ in $|\eta|<2.5$\\
 0.5--10 GeV,$|\eta|<2.5$& 0.5--5 & $\nch$ in $2.5<|\eta|<3.2$\\
 1--10 GeV,$|\eta|<2.5$  & 1--5 & $\nch$ in $3.2<|\eta|<4$\\
                         &        & $\nch$ in $4<|\eta|<5$  \\
                         &        & $\npart$\\\hline
\end{tabular}
\vspace*{-0.2cm}
\caption{Systematic variations in the analysis: $\pT$ ranges for defining $n(\pT)$ or $N(\pT)$, reference ranges for calculating $[\pT]$, and event activity variables for centrality classification.}
\vspace*{-0.2cm}
\label{tab:1}
\end{table}

For each subevent, we calculate the mean transverse momentum $[\pT]$, the unnormalized spectrum $N(\pT)$, and the fractional spectrum $n(\pT)$. Two-particle observables are evaluated in narrow bins of the event activity estimators to minimize detector and acceptance effects. Following the standard two-subevent methodology~\cite{ATLAS:2025ztg}, the $[\pT]$ fluctuations are measured as $\lr{(\delta[\pT])^2}= \lr{\delta[\pT]_a\delta[\pT]_b}$, and the covariance between spectral and $[\pT]$ fluctuations as \mbox{$\lr{\delta n(\pT) \delta [\pT]} =\frac{1}{2}\lr{\delta n_a(\pT) \delta [\pT]_b+\delta n_b(\pT) \delta [\pT]_a}$}. The offset term in Eqs.~\eqref{eq:6} and \eqref{eq:7a} is estimated using:
\begin{align}\label{eq:8}
\Delta v_0 = \frac{\lr{\delta N_a \delta[\pT]_b}}{\lr{N_a}\sqrt{\delta[\pT]_a\delta[\pT]_b}}
\end{align}

The independent source picture underlying HIJING provides useful scaling relations that serve as internal consistency checks. If particles used to define event activity are statistically independent of those used in the correlation analysis, the two-subevent quantities in Eq.~\eqref{eq:8} should exhibit simple power-law scaling with the number of sources $N_s$ (approximately proportional to $\npart$ in the HIJING framework):
\begin{align}\nonumber
&\lr{\delta N_a\delta N_b } \propto N_s  \\\nonumber
&\lr{\delta [\pT]_a \delta [\pT]_b } \propto 1/N_s  \\\nonumber
&\sqrt{\lr{\delta N_a\delta N_b}}/\lr{N_a} \propto 1/\sqrt{N_s}\\\nonumber
&\Delta v_0 \propto 1/\sqrt{N_s} \\\nonumber
&\lr{\delta N_a \delta [\pT]_b} \sim \mbox{const} \\\nonumber
&\sqrt{\lr{\delta N_a\delta N_b}}\sqrt{ \lr{\delta [\pT]_a \delta [\pT]_b}}\sim \mbox{const}\\\nonumber
&\rho(N_a,[\pT]_b) \equiv \frac{\lr{\delta N_a \delta [\pT]_b}}{\sqrt{\lr{\delta N_a\delta N_b }}\sqrt{\lr{\delta [\pT]_a \delta [\pT]_b }}}  \sim \mbox{const} \\\label{eq:9}
&\sqrt{N_a}\Delta v_0 \sim \mbox{const}
\end{align} 
Here, ``const'' denotes a value determined by particle production within individual sources (i.e., within each $p+p$ collision). These scaling relations provide important consistency checks for our analysis and help distinguish genuine correlations from statistical artifacts.

\begin{figure}[!h]
    \centering
    \includegraphics[width=0.9\linewidth]{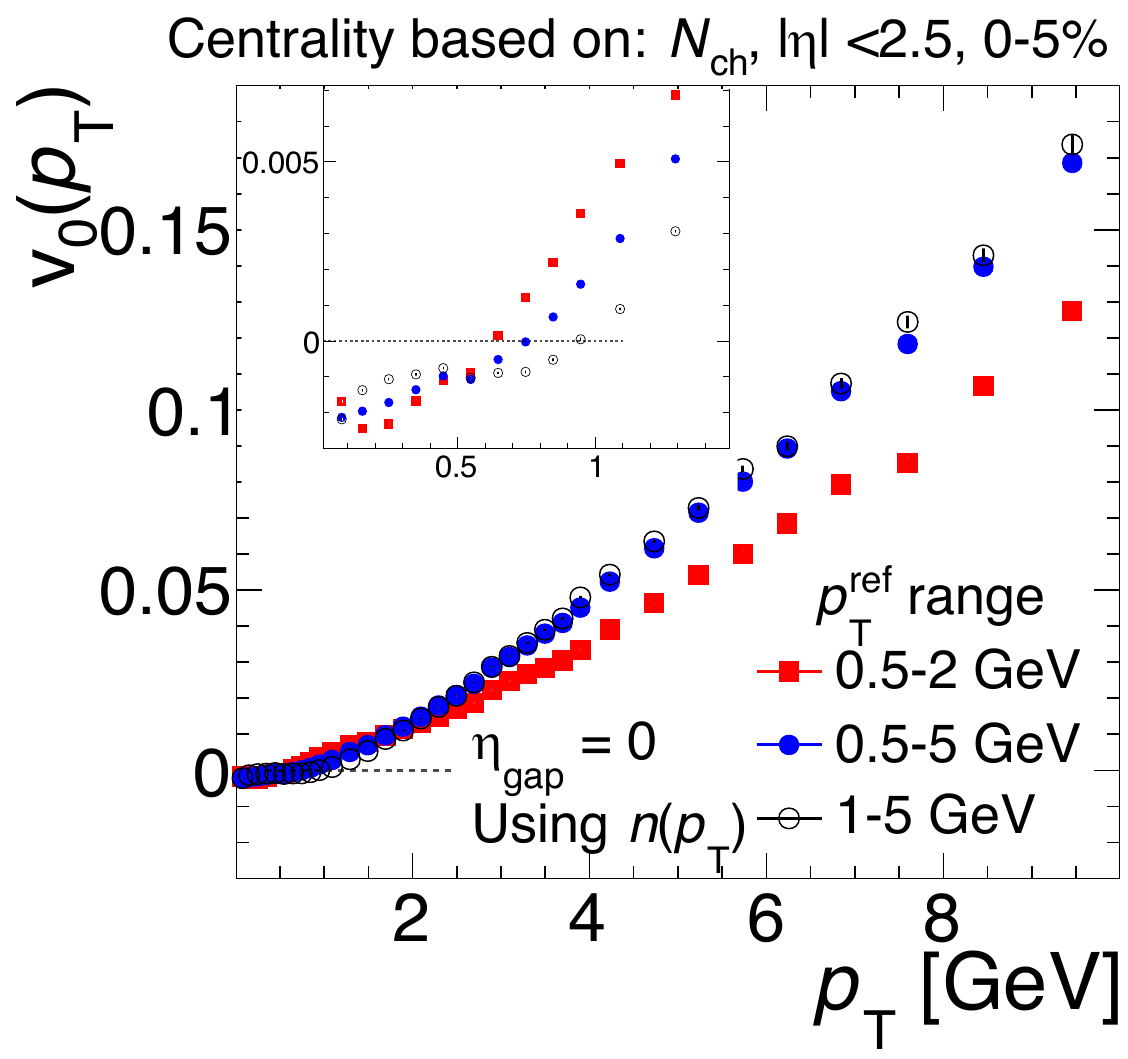}
\vspace*{-0.3cm}
    \caption{\textbf{Factorization behavior and impact of non-flow.} The $v_0(\pT)$ calculated for fractional spectra in 0--10 GeV range (Eq.~\eqref{eq:3}) for three $\pTref$ selections in 0--5\% most central Pb+Pb collisions. The inset shows a zoomed-in view of the low $\pT$ region. The $\pTref$-dependence reflects the influence of non-flow correlations.}
    \label{fig:2}
\end{figure}
\begin{figure*}[htbp]
    \centering
    \includegraphics[width=0.8\linewidth]{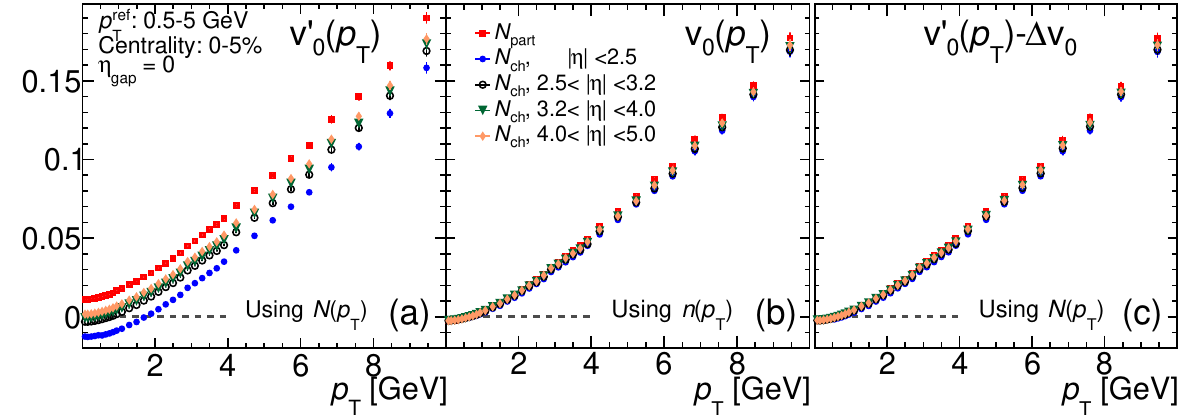}
    \caption{\textbf{The shift due to global multiplicity fluctuations via Eq.~\eqref{eq:2}.} The $v_0(\pT)$ obtained using $N(\pT)$ (left), $n(\pT)$ (middle), and $N(\pT)$ but applying the correction calculated via Eq.~\eqref{eq:6} in the two subevent format Eq.~\eqref{eq:8} (right) for several event activity selections in 0--5\% most central Pb+Pb collisions. They show that the spread due to global multiplicity fluctuations associated with different centrality estimators (left) can be largely removed by using normalized spectra (middle) and the difference between left and middle panel is well reproduced by Eq.~\eqref{eq:8}.}
    \label{fig:3}
\end{figure*}

{\bf Results.} \underline{\it Impact of non-flow correlations.} 
Figure~\ref{fig:2} presents $v_0(\pT)$ calculated using three different $\pTref$ selections in central collisions. The results reveal clear differences at high $\pT$ ($>$3~GeV), as expected from the dominance of non-flow correlations from jet production. These differences persist into the low-$\pT$ region (see inset), where they affect the zero-crossing point depending on the choice of $\pTref$. This is also a region where the resonance decays can contribute significantly. Importantly, the overall magnitude of $v_0(\pT)$ in HIJING is much smaller than that observed in experimental data~\cite{ATLAS:2025ztg}, confirming that collective radial flow---absent in HIJING---dominates the experimental measurements. 

\underline{\it Residual multiplicity fluctuations and offset.} 
The central result of this paper is illustrated in Fig.~\ref{fig:3}, which compares $v_0(\pT)$-like observables obtained using three different approaches: (a) using the unnormalized spectrum $N(\pT)$ via Eq.~\eqref{eq:5}, (b) using the normalized spectrum $n(\pT)$ via Eq.~\eqref{eq:3}, and (c) using $N(\pT)$ but applying the offset correction $\Delta v_0$ calculated from Eq.~\eqref{eq:8}. 

Figure~\ref{fig:3}a displays $v_0'(\pT)$ obtained from the unnormalized spectrum $N(\pT)$ for different event activity definitions. While all curves show a similar increasing trend with $\pT$, they exhibit different vertical offsets. These event-activity-dependent offsets can substantially alter the zero-crossing point $\pTc$: for event classes based on $\npart$ or $\nch$ measured in forward rapidity, $v_0'(\pT)$ never crosses zero, indicating the strong influence of global multiplicity fluctuations. In stark contrast, Fig.~\ref{fig:3}b shows that results based on the normalized spectrum $n(\pT)$ cluster together with a common zero-crossing point near the average transverse momentum of the inclusive spectrum ($\sim$1~GeV). This demonstrates that $v_0(\pT)$ obtained from fractional spectra is largely insensitive to the event activity definition, precisely as expected if it successfully isolates spectral shape fluctuations from global multiplicity effects. 

The key validation appears in Fig.~\ref{fig:3}c: when the offset correction $\Delta v_0$ via Eq.~\eqref{eq:6} is applied to the $N(\pT)$-based results, they collapse onto the $n(\pT)$-based results. This confirms that the difference between the two approaches is indeed a $\pT$-independent constant offset, exactly as predicted by our formalism.

\begin{figure}[!h]
    \centering
    \includegraphics[width=0.7\linewidth]{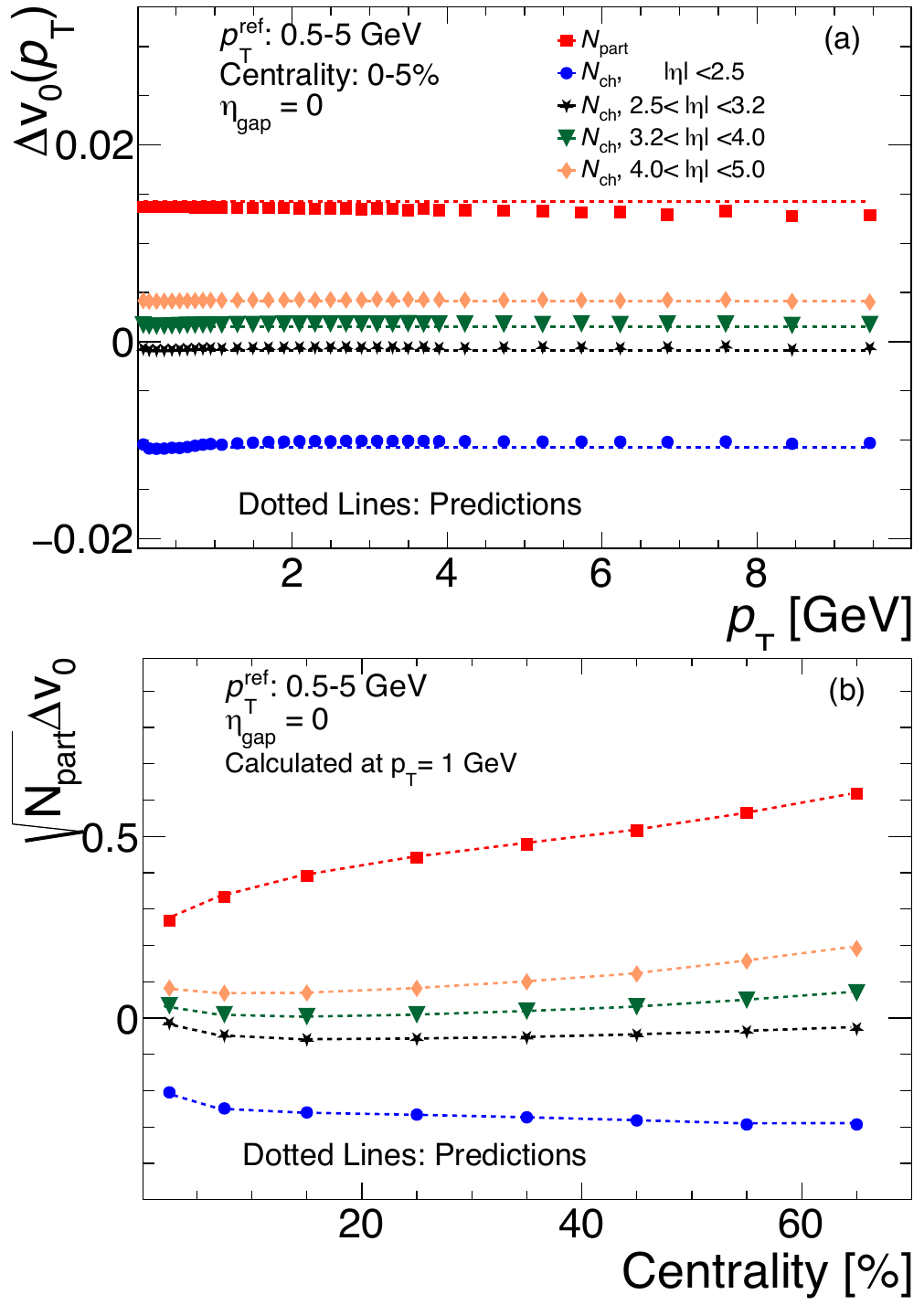}
    \caption{\textbf{The shift due to global multiplicity fluctuations via Eq.~\eqref{eq:2}.} Top: $v_0'(\pT)-v_0(\pT)$ (points) compared with $\Delta v_0$ estimated via Eq.~\eqref{eq:6} in its two-subevent format (dotted lines) in 0--5\% most central Pb+Pb collisions. Bottom: Centrality dependence of $\sqrt{\npart} (v_0'(\pT)-v_0(\pT))$ at $\pT=1$ GeV (points) compared with $\sqrt{\npart}\Delta v_0$ (dotted lines).}
    \label{fig:4}
\end{figure}

Further quantitative validation of the offset prediction is provided in Fig.~\ref{fig:4}. Figure~\ref{fig:4}a compares the measured difference $v_0'(\pT)-v_0(\pT)$ (symbols) with the predicted offset $\Delta v_0$ (dotted lines) as a function of $\pT$ for various event activity selections. The measured difference is remarkably flat across the entire $\pT$ range and is accurately reproduced by the simple analytical expression in Eq.~\eqref{eq:6}. Notably, the offset increases with the pseudorapidity separation between particles used for event activity classification and those used in the correlation analysis. This trend is consistent with the behavior of longitudinal multiplicity decorrelation, $\lr{\delta N(\eta_1)\delta N(\eta_2)}/[\lr{N(\eta_1)}\lr{N(\eta_2)}]$, measured previously by ATLAS~\cite{ATLAS:2016rbh}, and can be attributed to inherent forward-backward fluctuations in the initial state of heavy-ion collisions~\cite{Bzdak:2012tp,Jia:2015jga,Jia:2020tvb}.

Figure~\ref{fig:4}b extends this validation across centrality by plotting $\sqrt{\npart}[v_0'(\pT)-v_0(\pT)]$ at $\pT=1$~GeV compared with $\sqrt{\npart}\Delta v_0$. The excellent agreement between the two validates Eq.~\eqref{eq:6} across the full centrality range. Moreover, the approximately constant behavior of $\sqrt{\npart}\Delta v_0$ across centrality demonstrates excellent consistency with the independent source scaling expectation in Eq.~\eqref{eq:9}.

We note that for measurements performed in wide multiplicity bins, the correction term $\tau(\pT)$ in Eq.~\eqref{eq:5b} may not be small. In realistic heavy-ion data with collective flow, this term could in principle acquire a $\pT$-dependent component if multiplicity and spectral shape become dynamically correlated. However, this would manifest as a deviation from the flat behavior of $v_0'(\pT) - v_0(\pT)$, which can be checked empirically.

\begin{figure}[htbp]
    \centering
    \includegraphics[width=1\linewidth]{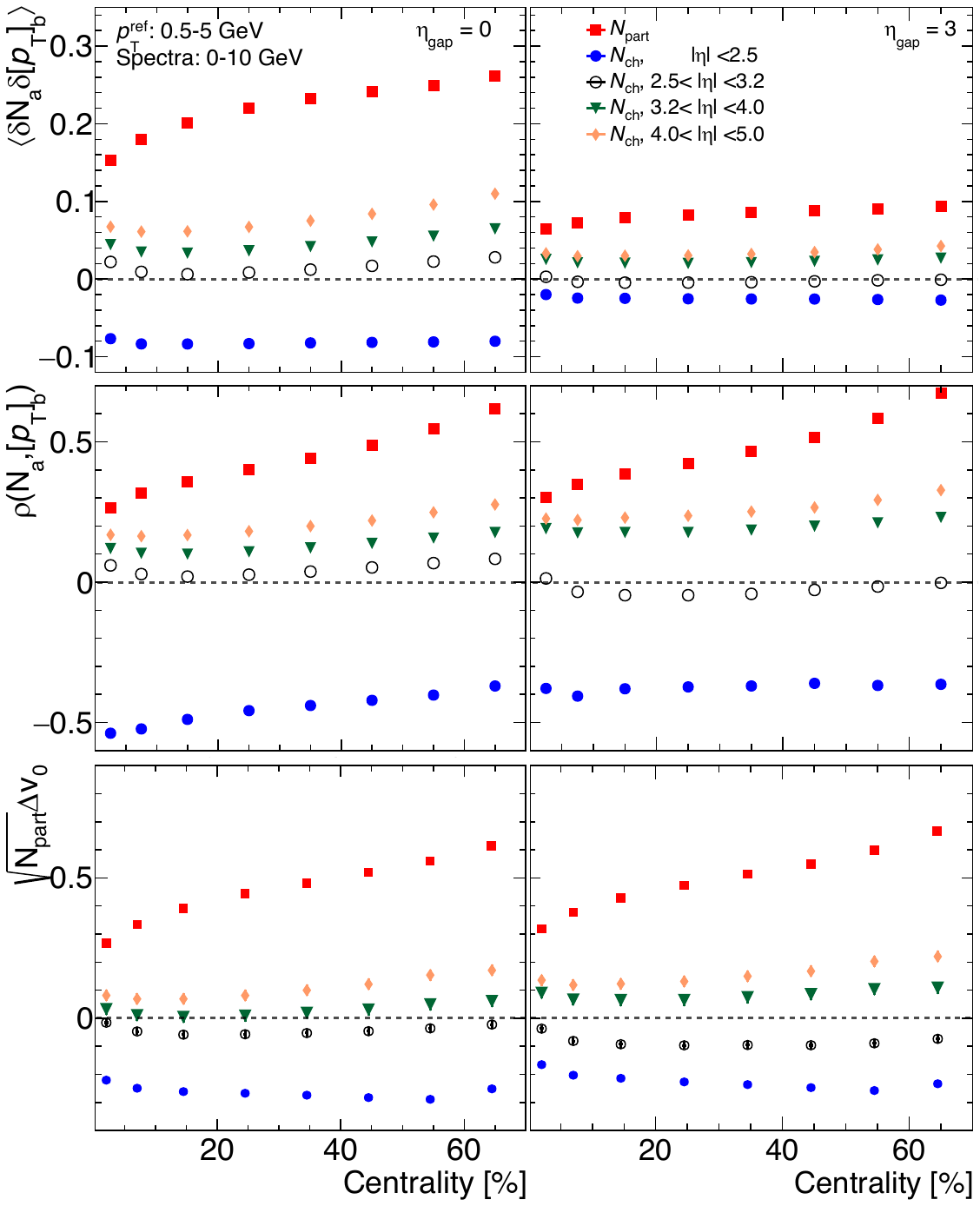}
    \caption{\textbf{Scaling behavior of components in $\Delta v_0$ in Eq.~\eqref{eq:8}:} $\lr{\delta N_a \delta [\pT]_b}$ (top row), $\rho(N_a, [\pT]_b)$ (middle row) and $\sqrt{N_a} \Delta v_0$ (bottom row) for $\eta_{\mathrm{gap}}=0$  (left column) and $\eta_{\mathrm{gap}}=3$ (right column), calculated in 0--5\% most central collisions, for various event activities classes.}
    \label{fig:5}
\end{figure}

\underline{\it Verification of independent source scaling.} 
To further validate the HIJING model's independent source picture and test the consistency of our framework, we examine the centrality dependence of the three key components entering $\Delta v_0$ in Eq.~\eqref{eq:8}: the covariance $\lr{\delta N_a \delta [\pT]_b}$, the correlation coefficient $\rho(N_a, [\pT]_b)$, and the scaled offset $\sqrt{N_a}\Delta v_0$. Figure~\ref{fig:5} displays these quantities for $\eta_{\mathrm{gap}}=0$ (left column) and $\eta_{\mathrm{gap}}=3$ (right column) across various event activity definitions.

According to the independent source scaling in Eq.~\eqref{eq:9}, these quantities should remain approximately constant across centrality when particles defining event activity are statistically independent of those used in the correlation analysis. Figure~\ref{fig:5} confirms this expectation for most configurations: the three components are either constant or vary slowly across centrality.

One configuration, however, requires special attention. When the event activity strongly overlaps with the analysis particles (e.g., $\nch$ in $|\eta|<2.5$, blue symbols), we observe negative values that deviate significantly from the general trend. This behavior arises from a statistical autocorrelation artifact. The total multiplicity $\nch$ comprises contributions from three regions: subevents $a$, $b$, and the middle region $c$, such that $\nch = N_a+N_b+N_c$. Requiring a fixed $\nch$ induces anticorrelations between $N_a$, $N_b$, and $N_c$, described by a multinomial distribution. When $\eta_{\mathrm{gap}}=0$, the middle region vanishes ($N_c=0$), and the anticorrelation between $N_a$ and $N_b$ or $[\pT]_b$ is strongest. As $\eta_{\mathrm{gap}}$ increases, more particles populate the middle region, weakening the anticorrelation between $N_a$ and $N_b$ (see bottom row of Fig.~\ref{fig:5}). Crucially, this autocorrelation artifact affects only $v_0'(\pT)$ calculated from the unnormalized spectrum $N(\pT)$, while $v_0(\pT)$ calculated from the normalized spectrum $n(\pT)$ remains unaffected by construction.

When particles defining event activity are separated from the analysis particles (no overlap), the correlations are close to zero or slightly positive. The correlation strength increases for $\nch$ measured in more forward $\eta$ regions and is strongest when using $\npart$ from the Glauber model. When $\npart$ is employed for event selection, the correlation shows a significant enhancement toward peripheral collisions. These positive correlations reflect genuine longitudinal dynamics encoded in the HIJING model rather than statistical artifacts. The stronger correlation for $\npart$-based event selection is expected since unlike multiplicity fluctuations, $\npart$ selection does not explicitly restrict the longitudinal dynamics inherent in particle production mechanisms (string, fragmentation, resonance etc).

\underline{\it Consequence of varying $\pT$ acceptance.} 
Finally, we address a practical experimental issue: different experiments employ different $\pT$ acceptance ranges. For example, ALICE typically measures $\pT>0.1$~GeV, CMS uses $\pT>0.3$~GeV, and ATLAS employs $\pT>0.5$~GeV in Pb+Pb collisions. The fractional spectra defined in these different ranges have distinct normalizations, leading to systematic differences in the extracted $v_0(\pT)$.

\begin{figure}[!h]
    \centering
    \includegraphics[width=1\linewidth]{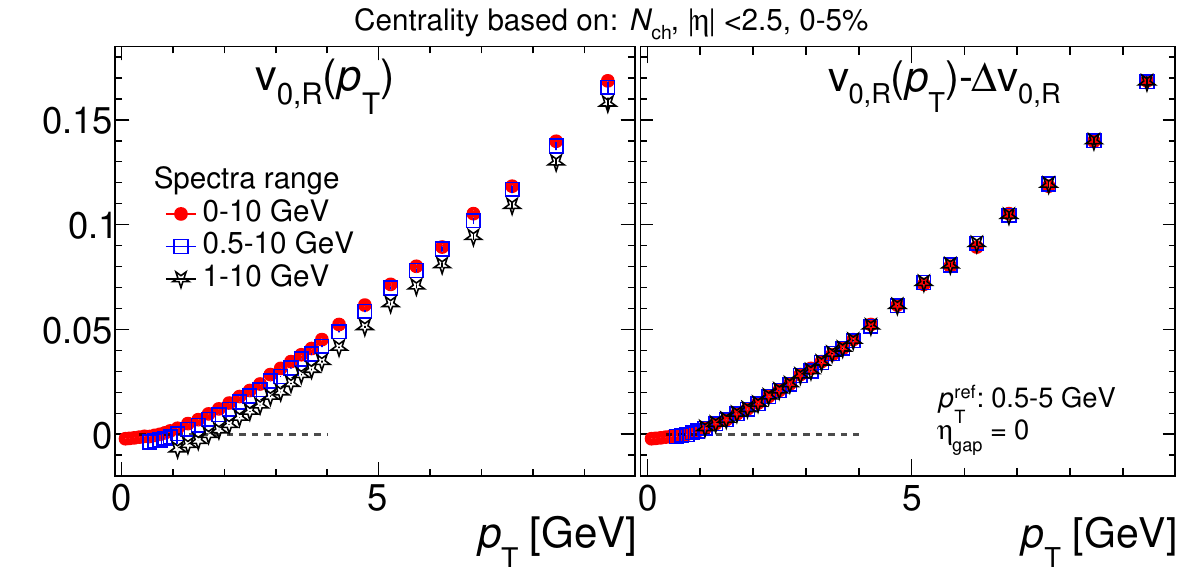}
    \caption{\textbf{Offset induced by the $\pT$-range choice for spectral normalization via Eq.~\eqref{eq:7aa}.} Left: The $v_0(\pT)$ obtained using spectra normalized using $N_R=\int_a^b N(\pT) d\pT$, obtained in three $\pT$ ranges, 0--10, 0.5--10 and 1--10 GeV, in 0-5\% most central collisions. Right: The results obtained after correcting for the offset predicted using Eq.~\eqref{eq:7a}. }
    \label{fig:6}
\end{figure}

According to Eq.~\eqref{eq:7c}, even when $N$ is fixed, the multiplicity $N_R$ within a restricted range still fluctuates due to binomial sampling. This residual fluctuation induces a downward shift in $v_0(\pT)$, moving the zero-crossing point from $\pTc\approx\lr{[\pT]}$ to $\pTc\approx \lr{[\pT]}_R$. Figure~\ref{fig:6}a confirms this expectation: measurements of $v_{0,R}(\pT)$ based on $n_R(\pT)$ show clear dependence on the $\pT$ range used for normalization, with zero-crossing points shifting systematically with the range choice. A nearly constant difference is indeed observed between ATLAS and ALICE measurements~\cite{Du:2025dpu}. However, Fig.~\ref{fig:6}b demonstrates that applying the offset correction from Eq.~\eqref{eq:7a} collapses these curves onto a common baseline, validating our prediction.

These results demonstrate that the location of the zero-crossing point of $v_0(\pT)$ is purely a kinematic effect determined by the average spectrum $\lr{n(\pT)}$ in the normalization range, and does not reflect genuine dynamical radial flow fluctuations. This has important implications for comparing $v_0(\pT)$ measurements across different experiments: direct comparison of absolute values is meaningless unless measurements are corrected to a common normalization range or, equivalently, vertically shifted to align their zero-crossing points.

{\bf Discussion and summary} 
Event-by-event fluctuations in the final particle spectrum $N(\pT)$ serve as a sensitive probe of collective radial flow in the quark-gluon plasma. Experimentally, these fluctuations are accessed through the observable $v_0(\pT)$ (or $v_0'(\pT)$), which characterizes the correlation between the normalized spectrum $n(\pT)$ (or the unnormalized spectrum $N(\pT)$) and the event-by-event spectral shape, encoded in the average transverse momentum $[\pT]$. The spectral fluctuations can be decomposed into contributions from global multiplicity fluctuations and local spectral shape fluctuations:
\[
\delta N(\pT) = \lr{n(\pT)} \delta N + \lr{N} \delta n(\pT)\;.
\]

We demonstrate that the separation between $\delta N$ and $\delta n(\pT)$ is fundamentally ambiguous. Using HIJING simulations, we systematically varied the event activity definition, the spectral normalization range, and the reference range for $[\pT]$ calculations. We found that $v_0'(\pT)$ based on the unnormalized spectrum $N(\pT)$ is sensitive to the event activity variable and affected by autocorrelation effects when centrality and analysis particles overlap. In contrast, $v_0(\pT)$ based on the normalized spectrum $n(\pT)$ is only sensitive to the normalization range. However, all formulations are related by a constant offset $\Delta v_0$ that reflects the correlation between $\delta N$ and $[\pT]$. Since the $[\pT]$ fluctuations are insensitive to these constant offsets, only the $\pT$-dependent variation of $v_0(\pT)$ contains meaningful physical information about collective radial flow. 

These findings have important practical implications. The zero-crossing point of $v_0(\pT)$ is influenced by volume or centrality fluctuations rather than reflecting genuine dynamical radial flow fluctuations. When comparing $v_0(\pT)$ across experiments or between data and theory, curves must be vertically shifted to align their zero-crossing points, or corrected to a common normalization range using Eq.~\eqref{eq:7a}. 

A practical universal convention would be to shift each $v_0(\pT)$ measurement vertically so that its zero crossing coincides with the species- and energy-specific $\lr{\pT}$ evaluated over the full $\pT$ range. This quantity is routinely published and provides a common, analysis-independent reference point. Direct comparison of absolute values without such corrections could be misleading, especially considering different experimental acceptances at the LHC (ALICE: $\pT>0.1$~GeV, CMS: $\pT>0.3$~GeV, ATLAS: $\pT>0.5$~GeV). 

However, to eliminate the normalization ambiguity entirely, we propose the derivative observable:
\begin{align}\label{eq:15}
v_{0s}(\pT)=dv_0(\pT)/d\pT\;.
\end{align}
According to Eq.~\eqref{eq:3}, $v_{0s}(\pT)$ directly quantifies the correlation between the local spectral slope $s(\pT)=dn(\pT)/d\pT /\lr{n(\pT)}\approx d\ln n(\pT)/d\pT$ and $[\pT]$:
\begin{align}\label{eq:16}
\frac{\lr{\delta s(\pT) \delta [\pT]}}{\lr{[\pT]}} = v_{0s}(\pT) v_{0,p}\;.
\end{align}
Since constant offsets vanish upon differentiation, $v_{0s}(\pT)$ is free from normalization ambiguities, providing an unambiguous alternative for characterizing radial flow fluctuations~\footnote{The measurement of $v_{0s}(\pT)$ naturally requires much more event statistics and careful handling of the $\pT$ bin width choices. A study of these experimental practicalities remains a subject for future investigation.}. Recent experimental results~\cite{ATLAS:2025ztg,ALICE:2025iud} imply that $v_{0s}(\pT)$ is positive at low $\pT$, crosses zero at $\pT\sim 3$--4~GeV, and approaches zero at higher $\pT$.

This work was supported by DOE Research Grant Number DE-SC0024602. 

\section*{Appendix}
\begin{figure}[!h]
\includegraphics[width=0.8\linewidth]{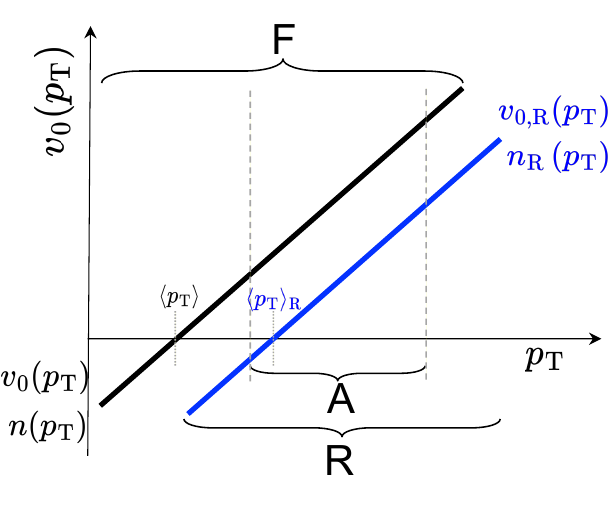}
\caption{\label{fig:a1} The $\pT$-differential spectra shape fluctuation calculated in the range $F$, $v_0(\pT)$, and range $R$, $v_{\mathrm{0,R}}(\pT)$. They differ by a constant vertical shift $s$, which is controlled by the difference between $\lr{\pT}$ and $\lr{\pT}_{R}$ (see main text). The range $A$ is the reference $\pT$ used to calculate the integral $v_0$ and for measuring $v_0(\pT)$ and $v_{\mathrm{ 0,R}}(\pT)$.}
\end{figure}

In this Appendix, we derive the sum rules relating the differential $v_0(\pT)$ to integrated $[\pT]$ fluctuations, and show how $v_0(\pT)$ measured in different $\pT$ ranges are related. We consider the full $\pT$ range ``$F$'', a restricted experimental range ``$R$'' (e.g., 0.5--10 GeV in ATLAS), and the reference $\pT$ range ``$A$'' used to calculate $[\pT]$ (see Fig.~\ref{fig:a1}). Quantities calculated in $F$ carry no subscript, while those in other ranges are labeled by subscript $R$ or $A$. We use $N(\pT)$ and $n(\pT)$ to denote the particle yield and fractional yield at $\pT$, and $N$ to denote the total multiplicity. The notation $\int_A$ represents integration of $\pT$ over range $A$. This discussion builds on Refs.~\cite{Schenke:2020uqq,Parida:2024ckk}.

\subsection{The observables in range $A$ expressed using variables in the full range}
The event-wise mean transverse momentum in $A$ and its ensemble average are given by
\small{\begin{align}\label{eq:a1}
[\pT]_A &= \frac{1}{N_A}\int_A \pT N(\pT)\\\nonumber
\lr{\pT}_A &\equiv \lr{[\pT]}_A =\lr{\frac{1}{N_A}\int_A \pT N(\pT)}\\\nonumber
&=\frac{1}{\lr{N_A}}\left[\int_A \pT \lr{N(\pT)} - \int_A \pT \lr{n_A(\pT)\delta N_A}\right]\\\label{eq:a2}
&\approx \frac{1}{\lr{N_A}}\int_A \pT \lr{N(\pT)}
\end{align}}\normalsize
The integral involving $\lr{n_A(\pT)\delta N_A}$ is a higher-order term corresponding to the correlated fluctuation of spectral shape and integrated multiplicity. The other term represents the per-particle average transverse momentum: if we calculate $\lr{\pT}$ on a per-track basis instead of an event-by-event basis, Eq.~\eqref{eq:a2} would be exact. From these, we obtain
\begin{align}\nonumber
\delta [\pT]_A &= \frac{1}{N_A}\int_A \pT \delta N(\pT) -\frac{\delta N_A}{N_A}\lr{\pT}_A \\\label{eq:a3}
&=\frac{1}{N_A}\int_A(\pT-\lr{\pT}_A)\delta N(\pT)
\end{align}
The appearance of $\lr{\pT}_A$ is critical because it absorbs the contribution from $\delta N_A$. Note that since $A$ is a subrange, $N_A$ still fluctuates even if the total multiplicity $N$ is fixed.

It is more convenient to express observables in terms of fractional spectra. Using $N(\pT) = n(\pT) N  = \delta n(\pT) N +\lr{n}N$, we have 
\begin{align}\nonumber
\lr{N(\pT)} &= \lr{N}\lr{n(\pT)}+ \lr{\delta N \delta n(\pT)}\\\label{eq:a4a}
\delta N(\pT) &= N\delta n(\pT) + \delta N n(\pT) - \lr{\delta N \delta n(\pT)}\;.
\end{align}
Eq.~\eqref{eq:a2} can be recast as,
\small{
\begin{align}\nonumber
\lr{\pT}_A &= \frac{\int_A \pT \lr{N(\pT)}}{\int_A \lr{N(\pT)}} = \frac{\lr{N}\int_A \pT \lr{n(\pT)} + \int_A \pT \lr{\delta N \delta n(\pT)} }{\lr{N}\int_A \lr{n(\pT)}+ \int_A \lr{\delta N \delta n(\pT)}}\\\nonumber
&\approx  \frac{\int_A \pT \lr{n(\pT)}}{\int_A \lr{n(\pT)}} \left(1+\frac{\int_A(\pT-\lr{\pT}_A)\lr{\delta N\delta n(\pT)}}{\lr{N_A} \lr{\pT}_A }\right)\\\label{eq:a4b}
&\approx  \frac{\int_A \pT \lr{n(\pT)}}{\int_A \lr{n(\pT)}}\;,
\end{align}}\normalsize
which implies 
\small{
\begin{align}\label{eq:a4c}
\int_A (\pT\!-\!\lr{\pT}_{\!A})\lr{n(\pT)}= \int_A \pT\!\lr{n(\pT)} - \lr{\pT}_{\!A}\!\int_A \!\lr{n(\pT)}=0
\end{align}}\normalsize
Plugging Eq.~\eqref{eq:a4a} into Eq.~\eqref{eq:a3}, we obtain
\begin{align}\nonumber
\delta [\pT]_A =&\frac{N}{N_A}\int_A(\pT-\lr{\pT}_A)\delta n(\pT) \\\nonumber
&+  \frac{\delta N }{N_A}\int_A (\pT-\lr{\pT}_A) \lr{n(\pT)} \\\nonumber
&- \frac{1}{N_A}\int_A (\pT-\lr{\pT}_A) \lr{\delta N n(\pT)}\\\label{eq:a4}
\approx & \frac{N}{N_A}\int_A(\pT-\lr{\pT}_A)\delta n(\pT)
\end{align}
The second term vanishes because of Eq.~\eqref{eq:a4c}, and the last higher-order term is also dropped. Again, the presence of $\lr{\pT}_A$ is essential to ensure the similarity of this equation to Eq.~\eqref{eq:a3}. Eqs.~\eqref{eq:a3} and \eqref{eq:a4} are valid independent of the underlying physics.

Following Ref.~\cite{Parida:2024ckk}, we assume that the EbE global radial flow fluctuation governs the EbE spectral fluctuation,
\begin{align}\label{eq:a5}
\frac{\delta n(\pT)}{\lr{n(\pT)}} = \frac{v_0(\pT)}{v_0}\frac{\delta [\pT]}{\lr{\pT}}
\end{align}
Plug this into Eq.~\eqref{eq:a4}, we obtain,
\small{
\begin{align}\nonumber
\frac{\delta [\pT]_A}{\lr{\pT}_A} &=\frac{N}{N_A\lr{\pT}_A}\int_A(\pT-\lr{\pT}_A)\delta n(\pT)\\\label{eq:a6}
&=\left[\!\frac{N}{N_A\!\lr{\pT}_A}\!\int_A\!(\pT\!-\!\lr{\pT}_A)\! \frac{v_0(\pT)}{v_0}\! \lr{n(\pT)}\!\right]\frac{\delta [\pT]}{\lr{\pT}}
\end{align}}\normalsize

Squaring this equation, averaging over events, and then using $\lr{(\delta [\pT])^2}/\lr{\pT}^2=v_0^2$, we obtain,
\[
\frac{\lr{(\delta [\pT]_A)^2}}{\lr{\pT}_A^2} =  \lr{\frac{N^2}{N_A^2}} \frac{\left(\int_A \!(\pT\!-\!\lr{\pT}\!_{A}\!)v_0(\pT)\! \lr{n(\pT)}\right)^2}{\lr{\pT}^2_A} 
\]
from this, and use the relation
\[
  \lr{\pT}_A = \frac{\lr{N}}{\lr{N_A}} \int_A \pT \lr{n(\pT)} + \frac{1}{\lr{N_A}} \int_A \pT \lr{\delta N \delta n(\pT)},
\]
but neglecting the $\lr{\delta N \delta n(\pT)}$ term, we obtain the sum rule,
\begin{align}\label{eq:a7}
v_{0,A}\equiv \frac{\sqrt{\lr{(\delta [\pT])_A^2}}}{\lr{\pT}_A}= f\times \frac{\int_A (\pT\!-\!\lr{\pT}_{\!A})v_0(\pT) \lr{n(\pT)}}{ \int_A \pT \lr{n(\pT)}},
\end{align}
where $f^2 =\frac{\lr{(N/N_A)^2}}{\lr{N}^2/\lr{N_A}^2}$. Assuming the fluctuation of $N_A$ comprises a component fully correlated with $N$ and a component at fixed $N$, then only the second component contributes, and we have $f\approx 1+ \frac{\lr{(\delta N_A)^2}}{2\lr{N_A}^2}|_N$. The deviation of $f$ from unity should be very small. 

This also yields the relation connecting the $v_{0}$ values calculated in two $\pT$ ranges, following Ref.~\cite{Parida:2024ckk}, $v_{0,A} = C_A v_0$, where 
\begin{align}\label{eq:a8}
C_A =  \frac{f \lr{N}}{\lr{N_A} \lr{\pT}_A} \int_A (\pT\!-\!\lr{\pT}_{\!A})\frac{v_0(\pT)}{v_0} \lr{n(\pT)}
\end{align}

Note that, because of the condition in Eq.~\eqref{eq:a4c}, Eqs.~\eqref{eq:a7} and \eqref{eq:a8} remain valid under a constant shift \mbox{$v_0(\pT)\rightarrow v_0(\pT)+c$}.
\\

\subsection{Relating the $v_0(\pT)$ measured in two $\pT$ ranges}
In principle, the normalization of the fractional spectra can be defined in any $\pT$ range. However, varying the choice of $\pT$ range effectively changes the nature of global multiplicity fluctuations defined in that range, as discussed in the main text. This results in a vertical shift that adjusts the zero-crossing point of $v_0(\pT)$ in each range to its own $\lr{\pT}$ value. In particular, the $v_0(\pT)$ calculated in ranges $F$ and $R$ are related by a constant:
\begin{align}
v_{\mathrm {0,R}}(\pT) = v_0(\pT) -s
\end{align}
This change in the zero-crossing point ensures that the corresponding weighted sum rule is satisfied.
\begin{align}
\int_F v_0(\pT) \lr{n(\pT)}=0, \int_R v_{\mathrm{ 0,R}}(\pT) \lr{n_{\mathrm{R}}(\pT)} =0
\end{align}

However, we would like to point out this sum rule is a direct consequence of the constraint $\int_F n(\pT) = 1$ and $\int_R n_R(\pT) = 1$. Different normalization ranges produce different sum rules and hence different zero-crossing points, none of which carries unique physical content. The resulting offsets are energy-dependent and particle-species-dependent, but they reflect volume or centrality fluctuations imposed by the chosen centrality estimator rather than genuine dynamical radial-flow information.

Using $\lr{n_{R} (\pT)} = \lr{N}/\lr{N_R} \lr{n (\pT)}$, Eq.~\eqref{eq:a7} is satisfied in both $F$ and $R$, 
\begin{align}\nonumber
v_{0,A}&\equiv \frac{\int_A (\pT\!-\!\lr{\pT}_{\!A})v_{\rm 0, R} (\pT) \lr{n_{\mathrm R}(\pT)}}{ \int_A \pT \lr{n_{R} (\pT)}}\\\nonumber
 &= \frac{\int_A (\pT\!-\!\lr{\pT}_{\!A})(v_{0}(\pT) -s) \lr{n(\pT)}}{ \int_A \pT \lr{n(\pT)}}\\\label{eq:a10}
 &= \frac{\int_A (\pT\!-\!\lr{\pT}_{\!A})v_{0}(\pT) \lr{n(\pT)}}{ \int_A \pT \lr{n(\pT)}}\;.
\end{align}

The insensitivity of $[\pT]$ fluctuations to the offset in $v_0(\pT)$ strengthens our conclusion that only the $\pT$-dependent variation of $v_0(\pT)$ is related to collective radial flow, because its baseline value is subject to a shift dictated by global multiplicity fluctuations.

Finally, we note that using $n(\pT)$ is more convenient than using $N(\pT)$, since for $N(\pT)$ Eq.~\eqref{eq:a5} must account for additional multiplicity fluctuations:
\begin{align}
\frac{\delta N(\pT)}{\lr{N(\pT)}} \approx \frac{\delta N }{\lr{N}}+\frac{v_0(\pT)}{v_0}\frac{\delta [\pT]}{\lr{\pT}}
\end{align}
which complicates the expression of Eqs.~\eqref{eq:a7} and \eqref{eq:a8} in terms of $v'_0(\pT)$.
\bibliography{../../v0pt}{}

\begin{thebibliography}{23}%
\makeatletter
\providecommand \@ifxundefined [1]{%
 \@ifx{#1\undefined}
}%
\providecommand \@ifnum [1]{%
 \ifnum #1\expandafter \@firstoftwo
 \else \expandafter \@secondoftwo
 \fi
}%
\providecommand \@ifx [1]{%
 \ifx #1\expandafter \@firstoftwo
 \else \expandafter \@secondoftwo
 \fi
}%
\providecommand \natexlab [1]{#1}%
\providecommand \enquote  [1]{``#1''}%
\providecommand \bibnamefont  [1]{#1}%
\providecommand \bibfnamefont [1]{#1}%
\providecommand \citenamefont [1]{#1}%
\providecommand \href@noop [0]{\@secondoftwo}%
\providecommand \href [0]{\begingroup \@sanitize@url \@href}%
\providecommand \@href[1]{\@@startlink{#1}\@@href}%
\providecommand \@@href[1]{\endgroup#1\@@endlink}%
\providecommand \@sanitize@url [0]{\catcode `\\12\catcode `\$12\catcode
  `\&12\catcode `\#12\catcode `\^12\catcode `\_12\catcode `\%12\relax}%
\providecommand \@@startlink[1]{}%
\providecommand \@@endlink[0]{}%
\providecommand \url  [0]{\begingroup\@sanitize@url \@url }%
\providecommand \@url [1]{\endgroup\@href {#1}{\urlprefix }}%
\providecommand \urlprefix  [0]{URL }%
\providecommand \Eprint [0]{\href }%
\providecommand \doibase [0]{http://dx.doi.org/}%
\providecommand \selectlanguage [0]{\@gobble}%
\providecommand \bibinfo  [0]{\@secondoftwo}%
\providecommand \bibfield  [0]{\@secondoftwo}%
\providecommand \translation [1]{[#1]}%
\providecommand \BibitemOpen [0]{}%
\providecommand \bibitemStop [0]{}%
\providecommand \bibitemNoStop [0]{.\EOS\space}%
\providecommand \EOS [0]{\spacefactor3000\relax}%
\providecommand \BibitemShut  [1]{\csname bibitem#1\endcsname}%
\let\auto@bib@innerbib\@empty
\bibitem [{\citenamefont {Heinz}\ and\ \citenamefont
  {Schenke}(2024)}]{Heinz:2024jwu}%
  \BibitemOpen
  \bibfield  {author} {\bibinfo {author} {\bibfnamefont {U.}~\bibnamefont
  {Heinz}}\ and\ \bibinfo {author} {\bibfnamefont {B.}~\bibnamefont
  {Schenke}},\ }\href@noop {} {\  (\bibinfo {year} {2024})},\ \Eprint
  {http://arxiv.org/abs/2412.19393} {arXiv:2412.19393 [nucl-th]} \BibitemShut
  {NoStop}%
\bibitem [{\citenamefont {Bozek}\ and\ \citenamefont
  {Broniowski}(2012)}]{Bozek:2012fw}%
  \BibitemOpen
  \bibfield  {author} {\bibinfo {author} {\bibfnamefont {P.}~\bibnamefont
  {Bozek}}\ and\ \bibinfo {author} {\bibfnamefont {W.}~\bibnamefont
  {Broniowski}},\ }\href {\doibase 10.1103/PhysRevC.85.044910} {\bibfield
  {journal} {\bibinfo  {journal} {Phys. Rev. C}\ }\textbf {\bibinfo {volume}
  {85}},\ \bibinfo {pages} {044910} (\bibinfo {year} {2012})},\ \Eprint
  {http://arxiv.org/abs/1203.1810} {arXiv:1203.1810 [nucl-th]} \BibitemShut
  {NoStop}%
\bibitem [{\citenamefont {Samanta}\ \emph {et~al.}(2024)\citenamefont
  {Samanta}, \citenamefont {Bhatta}, \citenamefont {Jia}, \citenamefont
  {Luzum},\ and\ \citenamefont {Ollitrault}}]{Samanta:2023amp}%
  \BibitemOpen
  \bibfield  {author} {\bibinfo {author} {\bibfnamefont {R.}~\bibnamefont
  {Samanta}}, \bibinfo {author} {\bibfnamefont {S.}~\bibnamefont {Bhatta}},
  \bibinfo {author} {\bibfnamefont {J.}~\bibnamefont {Jia}}, \bibinfo {author}
  {\bibfnamefont {M.}~\bibnamefont {Luzum}}, \ and\ \bibinfo {author}
  {\bibfnamefont {J.-Y.}\ \bibnamefont {Ollitrault}},\ }\href {\doibase
  10.1103/PhysRevC.109.L051902} {\bibfield  {journal} {\bibinfo  {journal}
  {Phys. Rev. C}\ }\textbf {\bibinfo {volume} {109}},\ \bibinfo {pages}
  {L051902} (\bibinfo {year} {2024})},\ \Eprint
  {http://arxiv.org/abs/2303.15323} {arXiv:2303.15323 [nucl-th]} \BibitemShut
  {NoStop}%
\bibitem [{\citenamefont {Heinz}\ and\ \citenamefont
  {Snellings}(2013)}]{Heinz:2013th}%
  \BibitemOpen
  \bibfield  {author} {\bibinfo {author} {\bibfnamefont {U.}~\bibnamefont
  {Heinz}}\ and\ \bibinfo {author} {\bibfnamefont {R.}~\bibnamefont
  {Snellings}},\ }\href {\doibase 10.1146/annurev-nucl-102212-170540}
  {\bibfield  {journal} {\bibinfo  {journal} {Ann. Rev. Nucl. Part. Sci.}\
  }\textbf {\bibinfo {volume} {63}},\ \bibinfo {pages} {123} (\bibinfo {year}
  {2013})},\ \Eprint {http://arxiv.org/abs/1301.2826} {arXiv:1301.2826
  [nucl-th]} \BibitemShut {NoStop}%
\bibitem [{\citenamefont {Hayrapetyan}\ \emph {et~al.}(2024)\citenamefont
  {Hayrapetyan} \emph {et~al.}}]{CMS:2024sgx}%
  \BibitemOpen
  \bibfield  {author} {\bibinfo {author} {\bibfnamefont {A.}~\bibnamefont
  {Hayrapetyan}} \emph {et~al.} (\bibinfo {collaboration} {CMS}),\ }\href
  {\doibase 10.1088/1361-6633/ad4b9b} {\bibfield  {journal} {\bibinfo
  {journal} {Rept. Prog. Phys.}\ }\textbf {\bibinfo {volume} {87}},\ \bibinfo
  {pages} {077801} (\bibinfo {year} {2024})},\ \Eprint
  {http://arxiv.org/abs/2401.06896} {arXiv:2401.06896 [nucl-ex]} \BibitemShut
  {NoStop}%
\bibitem [{\citenamefont {{ATLAS Collaboration}}(2024)}]{ATLAS:2024jvf}%
  \BibitemOpen
  \bibfield  {author} {\bibinfo {author} {\bibnamefont {{ATLAS
  Collaboration}}},\ }\href {\doibase 10.1103/PhysRevLett.133.252301}
  {\bibfield  {journal} {\bibinfo  {journal} {Phys. Rev. Lett.}\ }\textbf
  {\bibinfo {volume} {133}},\ \bibinfo {pages} {252301} (\bibinfo {year}
  {2024})},\ \Eprint {http://arxiv.org/abs/2407.06413} {arXiv:2407.06413
  [nucl-ex]} \BibitemShut {NoStop}%
\bibitem [{\citenamefont {Appelsh\"auser}\ \emph {et~al.}(1999)\citenamefont
  {Appelsh\"auser} \emph {et~al.}}]{NA49:1999inh}%
  \BibitemOpen
  \bibfield  {author} {\bibinfo {author} {\bibfnamefont {H.}~\bibnamefont
  {Appelsh\"auser}} \emph {et~al.} (\bibinfo {collaboration} {NA49}),\ }\href
  {\doibase 10.1016/S0370-2693(99)00673-5} {\bibfield  {journal} {\bibinfo
  {journal} {Phys. Lett. B}\ }\textbf {\bibinfo {volume} {459}},\ \bibinfo
  {pages} {679} (\bibinfo {year} {1999})},\ \Eprint
  {http://arxiv.org/abs/hep-ex/9904014} {arXiv:hep-ex/9904014} \BibitemShut
  {NoStop}%
\bibitem [{\citenamefont {{ALICE Collaboration}}(2014)}]{ALICE:2014gvd}%
  \BibitemOpen
  \bibfield  {author} {\bibinfo {author} {\bibnamefont {{ALICE
  Collaboration}}},\ }\href {\doibase 10.1140/epjc/s10052-014-3077-y}
  {\bibfield  {journal} {\bibinfo  {journal} {Eur. Phys. J. C}\ }\textbf
  {\bibinfo {volume} {74}},\ \bibinfo {pages} {3077} (\bibinfo {year}
  {2014})},\ \Eprint {http://arxiv.org/abs/1407.5530} {arXiv:1407.5530
  [nucl-ex]} \BibitemShut {NoStop}%
\bibitem [{\citenamefont {Adam}\ \emph {et~al.}(2019)\citenamefont {Adam} \emph
  {et~al.}}]{Adam:2019rsf}%
  \BibitemOpen
  \bibfield  {author} {\bibinfo {author} {\bibfnamefont {J.}~\bibnamefont
  {Adam}} \emph {et~al.} (\bibinfo {collaboration} {STAR}),\ }\href {\doibase
  10.1103/PhysRevC.99.044918} {\bibfield  {journal} {\bibinfo  {journal} {Phys.
  Rev. C}\ }\textbf {\bibinfo {volume} {99}},\ \bibinfo {pages} {044918}
  (\bibinfo {year} {2019})},\ \Eprint {http://arxiv.org/abs/1901.00837}
  {arXiv:1901.00837 [nucl-ex]} \BibitemShut {NoStop}%
\bibitem [{\citenamefont {Acharya}\ \emph {et~al.}(2024)\citenamefont {Acharya}
  \emph {et~al.}}]{ALICE:2023tej}%
  \BibitemOpen
  \bibfield  {author} {\bibinfo {author} {\bibfnamefont {S.}~\bibnamefont
  {Acharya}} \emph {et~al.} (\bibinfo {collaboration} {ALICE}),\ }\href
  {\doibase 10.1016/j.physletb.2024.138541} {\bibfield  {journal} {\bibinfo
  {journal} {Phys. Lett. B}\ }\textbf {\bibinfo {volume} {850}},\ \bibinfo
  {pages} {138541} (\bibinfo {year} {2024})},\ \Eprint
  {http://arxiv.org/abs/2308.16217} {arXiv:2308.16217 [nucl-ex]} \BibitemShut
  {NoStop}%
\bibitem [{\citenamefont {Abdulhamid}\ \emph {et~al.}(2024)\citenamefont
  {Abdulhamid} \emph {et~al.}}]{STAR:2024wgy}%
  \BibitemOpen
  \bibfield  {author} {\bibinfo {author} {\bibfnamefont {M.~I.}\ \bibnamefont
  {Abdulhamid}} \emph {et~al.} (\bibinfo {collaboration} {STAR}),\ }\href
  {\doibase 10.1038/s41586-024-08097-2} {\bibfield  {journal} {\bibinfo
  {journal} {Nature}\ }\textbf {\bibinfo {volume} {635}},\ \bibinfo {pages}
  {67} (\bibinfo {year} {2024})},\ \Eprint {http://arxiv.org/abs/2401.06625}
  {arXiv:2401.06625 [nucl-ex]} \BibitemShut {NoStop}%
\bibitem [{\citenamefont {Aad}\ \emph {et~al.}(2026)\citenamefont {Aad} \emph
  {et~al.}}]{ATLAS:2025ztg}%
  \BibitemOpen
  \bibfield  {author} {\bibinfo {author} {\bibfnamefont {G.}~\bibnamefont
  {Aad}} \emph {et~al.} (\bibinfo {collaboration} {ATLAS}),\ }\href {\doibase
  10.1103/ldcn-r2lq} {\bibfield  {journal} {\bibinfo  {journal} {Phys. Rev.
  Lett.}\ }\textbf {\bibinfo {volume} {136}},\ \bibinfo {pages} {032301}
  (\bibinfo {year} {2026})},\ \Eprint {http://arxiv.org/abs/2503.24125}
  {arXiv:2503.24125 [nucl-ex]} \BibitemShut {NoStop}%
\bibitem [{\citenamefont {Acharya}\ \emph {et~al.}(2026)\citenamefont {Acharya}
  \emph {et~al.}}]{ALICE:2025iud}%
  \BibitemOpen
  \bibfield  {author} {\bibinfo {author} {\bibfnamefont {S.}~\bibnamefont
  {Acharya}} \emph {et~al.} (\bibinfo {collaboration} {ALICE}),\ }\href
  {\doibase 10.1103/l36g-6f46} {\bibfield  {journal} {\bibinfo  {journal}
  {Phys. Rev. Lett.}\ }\textbf {\bibinfo {volume} {136}},\ \bibinfo {pages}
  {032302} (\bibinfo {year} {2026})},\ \Eprint
  {http://arxiv.org/abs/2504.04796} {arXiv:2504.04796 [nucl-ex]} \BibitemShut
  {NoStop}%
\bibitem [{\citenamefont {Parida}\ \emph {et~al.}(2024)\citenamefont {Parida},
  \citenamefont {Samanta},\ and\ \citenamefont {Ollitrault}}]{Parida:2024ckk}%
  \BibitemOpen
  \bibfield  {author} {\bibinfo {author} {\bibfnamefont {T.}~\bibnamefont
  {Parida}}, \bibinfo {author} {\bibfnamefont {R.}~\bibnamefont {Samanta}}, \
  and\ \bibinfo {author} {\bibfnamefont {J.-Y.}\ \bibnamefont {Ollitrault}},\
  }\href {\doibase 10.1016/j.physletb.2024.138985} {\bibfield  {journal}
  {\bibinfo  {journal} {Phys. Lett. B}\ }\textbf {\bibinfo {volume} {857}},\
  \bibinfo {pages} {138985} (\bibinfo {year} {2024})},\ \Eprint
  {http://arxiv.org/abs/2407.17313} {arXiv:2407.17313 [nucl-th]} \BibitemShut
  {NoStop}%
\bibitem [{\citenamefont {Mazeliauskas}\ and\ \citenamefont
  {Teaney}(2016)}]{Mazeliauskas:2015efa}%
  \BibitemOpen
  \bibfield  {author} {\bibinfo {author} {\bibfnamefont {A.}~\bibnamefont
  {Mazeliauskas}}\ and\ \bibinfo {author} {\bibfnamefont {D.}~\bibnamefont
  {Teaney}},\ }\href {\doibase 10.1103/PhysRevC.93.024913} {\bibfield
  {journal} {\bibinfo  {journal} {Phys. Rev. C}\ }\textbf {\bibinfo {volume}
  {93}},\ \bibinfo {pages} {024913} (\bibinfo {year} {2016})},\ \Eprint
  {http://arxiv.org/abs/1509.07492} {arXiv:1509.07492 [nucl-th]} \BibitemShut
  {NoStop}%
\bibitem [{\citenamefont {Gardim}\ \emph {et~al.}(2021)\citenamefont {Gardim},
  \citenamefont {Grassi}, \citenamefont {Ishida}, \citenamefont {Luzum},\ and\
  \citenamefont {Ollitrault}}]{Gardim:2020fxx}%
  \BibitemOpen
  \bibfield  {author} {\bibinfo {author} {\bibfnamefont {F.~G.}\ \bibnamefont
  {Gardim}}, \bibinfo {author} {\bibfnamefont {F.}~\bibnamefont {Grassi}},
  \bibinfo {author} {\bibfnamefont {P.}~\bibnamefont {Ishida}}, \bibinfo
  {author} {\bibfnamefont {M.}~\bibnamefont {Luzum}}, \ and\ \bibinfo {author}
  {\bibfnamefont {J.-Y.}\ \bibnamefont {Ollitrault}},\ }\href {\doibase
  10.1016/j.nuclphysa.2020.121892} {\bibfield  {journal} {\bibinfo  {journal}
  {Nucl. Phys. A}\ }\textbf {\bibinfo {volume} {1005}},\ \bibinfo {pages}
  {121892} (\bibinfo {year} {2021})},\ \Eprint
  {http://arxiv.org/abs/2002.01747} {arXiv:2002.01747 [nucl-th]} \BibitemShut
  {NoStop}%
\bibitem [{\citenamefont {Schenke}\ \emph {et~al.}(2020)\citenamefont
  {Schenke}, \citenamefont {Shen},\ and\ \citenamefont
  {Teaney}}]{Schenke:2020uqq}%
  \BibitemOpen
  \bibfield  {author} {\bibinfo {author} {\bibfnamefont {B.}~\bibnamefont
  {Schenke}}, \bibinfo {author} {\bibfnamefont {C.}~\bibnamefont {Shen}}, \
  and\ \bibinfo {author} {\bibfnamefont {D.}~\bibnamefont {Teaney}},\ }\href
  {\doibase 10.1103/PhysRevC.102.034905} {\bibfield  {journal} {\bibinfo
  {journal} {Phys. Rev. C}\ }\textbf {\bibinfo {volume} {102}},\ \bibinfo
  {pages} {034905} (\bibinfo {year} {2020})},\ \Eprint
  {http://arxiv.org/abs/2004.00690} {arXiv:2004.00690 [nucl-th]} \BibitemShut
  {NoStop}%
\bibitem [{\citenamefont {Gyulassy}\ and\ \citenamefont
  {Wang}(1994)}]{Gyulassy:1994ew}%
  \BibitemOpen
  \bibfield  {author} {\bibinfo {author} {\bibfnamefont {M.}~\bibnamefont
  {Gyulassy}}\ and\ \bibinfo {author} {\bibfnamefont {X.-N.}\ \bibnamefont
  {Wang}},\ }\href {\doibase 10.1016/0010-4655(94)90057-4} {\bibfield
  {journal} {\bibinfo  {journal} {Comput. Phys. Commun.}\ }\textbf {\bibinfo
  {volume} {83}},\ \bibinfo {pages} {307} (\bibinfo {year} {1994})},\ \Eprint
  {http://arxiv.org/abs/nucl-th/9502021} {arXiv:nucl-th/9502021} \BibitemShut
  {NoStop}%
\bibitem [{\citenamefont {Aaboud}\ \emph {et~al.}(2017)\citenamefont {Aaboud}
  \emph {et~al.}}]{ATLAS:2016rbh}%
  \BibitemOpen
  \bibfield  {author} {\bibinfo {author} {\bibfnamefont {M.}~\bibnamefont
  {Aaboud}} \emph {et~al.} (\bibinfo {collaboration} {ATLAS}),\ }\href
  {\doibase 10.1103/PhysRevC.95.064914} {\bibfield  {journal} {\bibinfo
  {journal} {Phys. Rev. C}\ }\textbf {\bibinfo {volume} {95}},\ \bibinfo
  {pages} {064914} (\bibinfo {year} {2017})},\ \Eprint
  {http://arxiv.org/abs/1606.08170} {arXiv:1606.08170 [hep-ex]} \BibitemShut
  {NoStop}%
\bibitem [{\citenamefont {Bzdak}\ and\ \citenamefont
  {Teaney}(2013)}]{Bzdak:2012tp}%
  \BibitemOpen
  \bibfield  {author} {\bibinfo {author} {\bibfnamefont {A.}~\bibnamefont
  {Bzdak}}\ and\ \bibinfo {author} {\bibfnamefont {D.}~\bibnamefont {Teaney}},\
  }\href {\doibase 10.1103/PhysRevC.87.024906} {\bibfield  {journal} {\bibinfo
  {journal} {Phys. Rev. C}\ }\textbf {\bibinfo {volume} {87}},\ \bibinfo
  {pages} {024906} (\bibinfo {year} {2013})},\ \Eprint
  {http://arxiv.org/abs/1210.1965} {arXiv:1210.1965 [nucl-th]} \BibitemShut
  {NoStop}%
\bibitem [{\citenamefont {Jia}\ \emph {et~al.}(2016)\citenamefont {Jia},
  \citenamefont {Radhakrishnan},\ and\ \citenamefont {Zhou}}]{Jia:2015jga}%
  \BibitemOpen
  \bibfield  {author} {\bibinfo {author} {\bibfnamefont {J.}~\bibnamefont
  {Jia}}, \bibinfo {author} {\bibfnamefont {S.}~\bibnamefont {Radhakrishnan}},
  \ and\ \bibinfo {author} {\bibfnamefont {M.}~\bibnamefont {Zhou}},\ }\href
  {\doibase 10.1103/PhysRevC.93.044905} {\bibfield  {journal} {\bibinfo
  {journal} {Phys. Rev. C}\ }\textbf {\bibinfo {volume} {93}},\ \bibinfo
  {pages} {044905} (\bibinfo {year} {2016})},\ \Eprint
  {http://arxiv.org/abs/1506.03496} {arXiv:1506.03496 [nucl-th]} \BibitemShut
  {NoStop}%
\bibitem [{\citenamefont {Jia}\ \emph {et~al.}(2020)\citenamefont {Jia},
  \citenamefont {Zhang},\ and\ \citenamefont {Xu}}]{Jia:2020tvb}%
  \BibitemOpen
  \bibfield  {author} {\bibinfo {author} {\bibfnamefont {J.}~\bibnamefont
  {Jia}}, \bibinfo {author} {\bibfnamefont {C.}~\bibnamefont {Zhang}}, \ and\
  \bibinfo {author} {\bibfnamefont {J.}~\bibnamefont {Xu}},\ }\href {\doibase
  10.1103/PhysRevResearch.2.023319} {\bibfield  {journal} {\bibinfo  {journal}
  {Phys. Rev. Res.}\ }\textbf {\bibinfo {volume} {2}},\ \bibinfo {pages}
  {023319} (\bibinfo {year} {2020})},\ \Eprint
  {http://arxiv.org/abs/2001.08602} {arXiv:2001.08602 [nucl-th]} \BibitemShut
  {NoStop}%
\bibitem [{\citenamefont {Du}(2026)}]{Du:2025dpu}%
  \BibitemOpen
  \bibfield  {author} {\bibinfo {author} {\bibfnamefont {L.}~\bibnamefont
  {Du}},\ }\href {\doibase 10.1103/9pvq-ph1c} {\bibfield  {journal} {\bibinfo
  {journal} {Phys. Rev. C}\ }\textbf {\bibinfo {volume} {113}},\ \bibinfo
  {pages} {014901} (\bibinfo {year} {2026})},\ \Eprint
  {http://arxiv.org/abs/2508.07184} {arXiv:2508.07184 [nucl-ex]} \BibitemShut
  {NoStop}%
\end{thebibliography}%
\bibliographystyle{apsrev4-1}
\end{document}